\documentclass[twocolumn,aps,prb,showpacs,amsmath,amssymb]{revtex4}

\usepackage{graphicx}
\usepackage{dcolumn}
\usepackage{bm}
\newcommand{\Ang}{\,\mathrm{\AA}}
\newcommand{\tr}{\mathrm{Tr}}
\newcommand{\ev}{\,\mathrm{eV}}

\begin{document}
{\parbox[b]{1in}{\hbox{\tt INT-PUB-08-47}}}
\title{Theory of the spontaneous buckling of doped graphene}
\author{Doron Gazit}
\email{doron.gazit@mail.huji.ac.il}
\affiliation{Institute for Nuclear Theory, University of Washington, 
Box 351550, Seattle, Washington 98195, USA}

\date{\today}

\begin{abstract}
Graphene is a realization of an esoteric class of materials -- electronic crystalline membranes.
We study the interplay between the free electrons and the two-dimensional crystal, and find that it induces a substantial effect on the elastic structure of the membrane. For the hole-doped membrane, in particular, we predict a spontaneous buckling. In addition, attenuation of elastic waves is expected, due to the effect of corrugations on the bulk modulus. These discoveries have a considerable magnitude in graphene, affecting both its mesoscopic structure, and its electrical resistivity, which has an inherent asymmetry between hole- and electron-doped graphene.
\end{abstract}

\pacs{73.22.-f  68.35.Iv  73.50.-h  73.63.-b.}

\maketitle
\begin{section}{INTRODUCTION}
The isolation of graphene \cite{2004SciNovoselov} has not only
provided the catalyst for countless discoveries \cite{2007NatMatGeim,2007RMParXiv}, but also heralded the unification of two independent branches of physics: the structure of crystalline membranes on one hand,\cite{Membranes_Book,2001PhRBowick} and the properties of quantum field theories (QFTs) of fermions in two dimensions (2D) on the other hand. \cite{1994PhysRevBLudwig, 2008PhysRevBChamon} In its structure, graphene is the ultimate crystalline membrane -- made of carbon atoms arranged in a 2D hexagonal lattice. Its low-energy electronic structure is analogous to a massless Dirac fermion, establishing graphene as a table-top experimental device for the study of 2D QFT.\cite{2006NatPhKatsnelson} Clearly, graphene offers an unprecedented opportunity  to examine the interplay between these aspects of the material.\\ 
The two-dimensional structure is known to be an important element in the electronic
transport properties, due to scattering of charge carriers off corrugations
\cite{2008RSPTAKatsnelson} and in-plane deformations
\cite{2008PhRvBHwang,2007PhRvBStauber}. However, is the surface structure affected by
the electronic properties? In this Brief Report we argue that the answer to this question is positive. We evaluate the contribution of the free-electrons to the elastic free energy, and find that the electron-phonon interactions lead to a substantial effect on the elastic constants. In particular, we find that the bulk modulus acquires a correction which depends on the wavelength of the excited elastic waves, thus implying an attenuation of these waves. In addition, and more important, we find that a mesoscopic appearance of ripples, i.e., buckling, is possible for a hole-doped membrane. To our knowledge, this is a unique property of electronic crystalline membranes in general, and graphene in particular.
Seeing that the scattering of charge-carriers off ripples increases the resistivity of the material, this prediction hints that the origin of the asymmetry found experimentally \cite{2008PhRvLBolotin,2008PhRvLMorozov} between the resistivity of hole- and electron-doped graphene, is this coupling of the physical structure and electronic characteristics. As such, this esoteric effect should be taken into account when designing graphene based technological applications.
\end{section}
\begin{section}{ELASTIC PROPERTIES OF GRAPHENE}
As a 2D material, the mere existence of graphene was in doubt, due to the
Mermin-Wagner theorem,\cite{1966PRLMerminWagner} which forbids the
existence of long-range order in two-dimensions. This seeming
contradiction can be resolved by introducing small
out-of-plane crumpling, which suppresses thermal vibrations.\cite{Membranes_Book,1987JPhNelson} Extensive theoretical, numerical, and
experimental investigations of crystalline membranes have established
the existence and stability of a low temperature phase, characterized
by corrugations and ripples, however asymptotically flat.
\cite{2001PhRBowick} The stability of graphene, even without the
support of a substrate, was found 
experimentally \cite{2007NatMeyer} and numerically.\cite{2007NatMaFasolino}

In order to model this ``almost-flat'' phase of a membrane, it is useful
to describe the deviation from the ideal phase, i.e., a flat surface with perfect lattice. In-plane
deformations are characterized by a two dimensional vector field
$\vec{u}$, and the out-of-plane deformation by a field $h$. This deformation changes the distance between two points on the surface, initially separated by an infinitesimal vector $d\vec{x}=(dx_1,dx_2)$, by $2\sum_{i,j=1}^2 u_{ij}dx_i dx_j$, where $u_{ij}$ is the strain tensor,
$u_{ij}=\frac{1}{2}\left(\partial_i u_j+\partial_j
u_i\right)+\frac{1}{2}(\partial_i h)(\partial_j h)$. 
The mesoscopic structure of the surface is determined by a free energy functional,
which has to preserve the homogeneity and isotropy of this limit -- the elastic free energy:
\begin{eqnarray} \label{Eq:Elastic_energy}
F[u,h] &=&\frac{1}{2}\int d^2\vec{x}\kappa (\Delta h)^2 + 
\\ \nonumber 
 &+& \frac{1}{2}\int
d^2\vec{x}\left[2\mu \sum_{i,j=1}^2 u_{ij}^2 +\lambda \left(\sum_{i=1}^{2}u_{ii}\right)^2 \right]. 
\end{eqnarray}
The coefficients are the elastic properties of the membrane, its bending energy $\kappa
\approx 1.1 \ev$, bulk modulus $\lambda+\mu \approx 11
\ev\Ang^{-2}$, and shear modulus $\mu \approx 9
\ev\Ang^{-2}$.\cite{2007NatMaFasolino,1972PhRvBNicklow}
They originate in the $\sigma$ bond between the carbon
atoms, which is a consequence of $sp^2$ hybridization, that forms a deep valence band. This fact is evident in the high creation energy for structural defects, reflected
experimentally in the high lattice quality. In fact, lattice defects were not
observed in graphene, even at strain values of about 1\%,\cite{2007NatMeyer,2007NatMaFasolino,2005NatZhang} validating the elastic approximation up to high strains.    
\end{section}
\begin{section}{ELECTRONIC STRUCTURE OF CORRUGATED GRAPHENE}
An additional electron in a $p$ orbital differentiates graphene from other crystalline membranes. This orbital, which is perpendicular to
the planar structure, forms a half filled $\pi$ band. This electron is responsible for one of the unusual aspects of graphene -- its celebrated Dirac type low-energy excitations. At
low-energy, the hexagonal lattice leads to a linearly vanishing
density of states around two points, named Dirac points, in the reciprocal
space.\cite{2005NatNovoselov} This dispersion relation characterizes massless, chiral Dirac fermions, with an effective speed of light $v_f \approx c/300$, whose size is dictated by the hopping integral of
the carbon bond. This picture leads to a ballistic movement of the electron. Experimentally this mobility is limited, probably due to the interaction of electron with phonon fields, i.e., the surface structure.\cite{2008PhRvBHwang, 2008PhRvLBolotin, 2007PhRvBStauber,2008PhRvLMorozov}

The dominant effect the complex mesoscopic structure of graphene has on the
ideal QED electronic structure arises from the deformation of the lattice.
The source of this deformation potential is the local change in the Fermi energy measured from the bottom of the $\sigma$ bond, proportional to the change in area $\delta S \propto \sum_{i=1}^2 u_{ii}$. The resulting deformation potential has the form $V_s=D(\sum_{i=1}^2 u_{ii})$ with, $D= 10-30 \ev$.\cite{2002PhysRevBAndo} We note, as stated by Hwang and Das-Sarma,\cite{2008PhRvBHwang} that since the source of this potential is not a Coulomb potential, its screening can be neglected.

Another effect of the deformation of the lattice is a change in the hoping integral, which manifests itself as an effective gauge field in the Dirac picture.\cite{2002PhysRevBAndo,2008PhRvBGuinea} Its structure is determined by the symmetry of the lattice, and its characteristic size is much smaller than the deformation energy, $g_2 \approx 1-4\ev$,\cite{2002PhysRevBAndo} thus we will postpone the discussion on its effect on the elastic free enegy to a different work.\cite{Gazit} We comment, however, that this pesudo magnetic field leads to preferred directions of ripples, following the underlying lattice symmetry.

The energetics of the $\pi$ electron, in the continuum limit, can be
summarized in a QED like action, of a fermion in non-zero chemical
potential. The Matsubara action of an electron, of inverse temperature $\beta$, is
\begin{eqnarray} \nonumber
S&=&-\sum_{n=1}^{N}
\int_0^\beta d\tau \int d^2 {\vec{x}} \bar{\Psi}_n (\tau,\vec{x}) \cdot \\ \nonumber &\cdot&\left ( \gamma^0(\partial_0+i(Du_{ii}+\delta V)) + v_f \vec{\gamma} \cdot \vec{\nabla} \right)\cdot \Psi_n(\tau,\vec{x}) .
\end{eqnarray}
Here, $n$ is the index due to the $N=2$ spin degeneracy, $\Psi$ is a four component Dirac fermion, reflecting the two sub-lattices and two Dirac cones. The $\gamma$ matrices satisfy the Clifford algebra, $\{\gamma_\mu,\gamma_\nu\}=2\delta_{\mu\nu}$. Note that the total electro-chemical potential $V$ is a sum of the deformation energy, and of any other external electro-chemical potential $\delta V$. As will be shown later, the interaction of the deformation energy with an external electro-chemical potential can lead to
buckling in graphene, as the external potential can compensate a deformation energy, allowing a deformed lattice. In this way, a buckled phase can be favored.
In order for this state to be the equilibrium phase, a mutual minimization of the elastic membrane energy
(Eq.~\ref{Eq:Elastic_energy}), and the electron energy is needed.\cite{2008EPLKim,2008PhRvBGuinea}
\end{section}
\begin{section}{CONTRIBUTION OF FREE ELECTRONS TO THE FREE ENERGY}
Prominent is the fact that the characteristic time scale of
the electrons is determined by $v_f \approx 10^8 \,\mathrm{cm/sec}$,  whereas the mesoscopic structure of graphene changes in a much longer time scale, dictated by the speed
of sound $v_{ph}\approx 2\cdot 10^6\,\mathrm{cm/sec}$. As a result, electronic excitations are substantially more energetic than the
phonons, or elastic, excitations, and thus the electronic degrees-of-freedom can be integrated out. 
In addition, we assume that the strength of the
electro-chemical potential is weak, keeping terms only to second order in the
strength $V^2$. The resulting contribution of the $\pi$ electrons to the free energy is
$ F_{electron}=\frac{1}{2}\int \frac{d^2\vec{q}}{(2\pi)^2}  \Pi^V(\vec{q};\beta) V_q^2 $, where the subscript $q$ denotes Fourier transform, i.e., for a constant external electro-chemical potential: $V_q=D(\sum_{i=1}^2 u_{ii})_q+(2\pi)^2 \delta V \delta(\vec{q})$. $\Pi^V(\vec{q};\beta)$ is the polarization operator of momentum $q$, which in the one--loop level is: 
\begin{eqnarray} \nonumber
\Pi^V( {q}^\mu) \equiv \frac{2}{\beta}\sum_{n=-\infty}^{n=\infty} \int\frac{d^2\vec{k}}{(2\pi)^2} \frac{k_\alpha(k+q)_\beta}{k^2({k}+{q})^2} \tr \gamma^0 \gamma^\alpha \gamma^0 \gamma^\beta, 
\end{eqnarray}
where the Matsubara frequencies are $k^0_n=\frac{2\pi}{\beta}(n+\frac{1}{2})$. The polarization operator of the static electro-chemical potential, i.e., $q^0=0$, is thus:\cite{2008PhRvBDillenschneider}
\begin{eqnarray} \nonumber
\Pi^V(\vec{q};\beta) =  \frac{4}{\pi(\hbar v_f)^2\beta} \int_0^1 dx \ln \left[ 2 \cosh\left( \frac{\hbar v_f\beta q}{2} \sqrt{x(1-x)}\right) \right].
\end{eqnarray}
For $\hbar v_f \beta q  \ll  1$, one can show that $\Pi^V(q;\beta) = \frac{4 \ln 2}{\pi (\hbar v_f)^2 \beta}$. In the opposite limit $\hbar v_f \beta q \gg  1$, $\Pi^V(q;\beta) =  \frac{q}{4\hbar v_f}$
The transition between these asymptotic behaviors occurs rather abruptly, for $\hbar v_f \beta q \approx 2$, as can be seen in Figure~\ref{Fig:polarization}.

\begin{figure}[t]
\rotatebox{0}{\resizebox{7.1cm}{!}{
\includegraphics[clip=true,viewport=3cm 15.5cm 20cm 25.5cm]{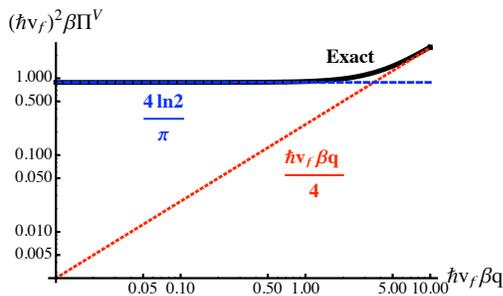}} }
\caption{(Color online) Dimensionless polarization operator $(\hbar v_f)^2 \beta \Pi^V$ as a function of the dimensionless momentum $\hbar v_f \beta q$. }
\label{Fig:polarization}
\end{figure}
\end{section}
\begin{section}{EFFECT OF FREE ELECTRONS ON THE BULK MODULUS} One can clearly see that the contribution to the free energy leads to a term in the free energy which is proportional to $|\sum_{i=1}^2 u_{ii}|^2$, i.e., increases the bulk modulus. Specifically, the Lame constant $\lambda$ is renormalized,  $\lambda(q) \equiv \lambda + D^2 \Pi^V(\vec{q};\beta)$. This implies a finite contribution even for $q\rightarrow 0$, of $0.6 \ev \Ang^{-2} \left(\frac{D}{30\ev}\right)^2 \left(\frac{T}{300^\circ{\mathrm{K}}}\right)$. For ripples on the order of $50-150 \Ang$, which are found experimentally on graphene \cite{2007NatMeyer,2007NatMaFasolino} and around room temperature, the appropriate limit is $\hbar v_f \beta q \gg 1$, which implies an addition of $3.85 {\ev \cdot \Ang^{-2}} \left(\frac{D}{30\ev}\right)^2 \left(\frac{q}{0.1\Ang^{-1}}\right)$. This contribution is substantial compared to the experimentally found $\lambda=2 \ev \Ang^{-2}$, and depends on the wave-number, thus should be observed as an attenuation of sound wave packets in graphene. 

The dependence of this attenuation on the deformation energy $D$ is quadratic. This strong dependence can be used to measure this parameter, whose value is known rather poorly. This parameter is of immense importance, as the resistivity induced by the deformation potential is considered to be the dominant source of carrier scattering at high carrier densities.\cite{2008PhRvLMorozov} $D$ is usually extracted from its contribution to resistivity, hence an independent measurement, as suggested here, is valuable.
\end{section}
\begin{section} {SPONTANEOUS BUCKLING} 
However, the most interesting effect stems from the term in $F_{electron}$ proportional to the cross-product of $\delta V_q$ and $\sum_{i=1}^2 u_{ii}$, which has the form $\delta F = \tau \int d^2{\vec{x}} \sum_{i=1}^2 u_{ii}$ with $\tau=D \Pi^V(\vec{q}=0;\beta) \delta V$.  This term leads to a buckling instability. The reason for that is easily seen when considering vanishing in-plane deformations, and keeping quadratic terms in $h$. In this case:
\begin{equation}
F[u,h]=\frac{1}{2}\int d^2\vec{x} \left[\kappa (\Delta h)^2 +\tau (\vec{\nabla} h)^2 \right].
\end{equation}
If $\tau<0$, this free-energy is minimized when the spatial configuration has a correlation length of
the order of $\xi = 2\pi \sqrt{\frac{\kappa}{|\tau|}}$.\cite{2008PhRvBGuinea,1988PhRvLGuitter} Intuitively, one may understand the condition $\tau<0$, by thinking about a piece
of paper, which is not buckled when stretched, but is buckled when
contracted. The same happens here, with the chemical potential, which takes the part of the contraction stress in the analogy to a paper. A negative potential pulls the ions, which are held by the elastic forces. The competition can be resolved by the ripple creation. A positive potential, however, results in local stretching, with no advantage in rippling.

The external potential $\delta V$, in the single electron picture, is reflected in the chemical potential of an electron. Sources of such chemical potential are numerous. For example, adsorption of molecules on graphene could provide a
simple possible mechanism to check the theory, predicting an inherent difference between exothermic and endothermic adsorptions.  For example, doping with $\text{NO}_2$ molecules, which behave as holes (acceptors), has been demonstrated to high doping levels, $n\propto10^{13} \text{cm}^{-2}$.\cite{2007NatMaSchedin,2008ZhouPhysRevB} Another interesting source of chemical potential is a finite density of surface charge carriers. A source of such charge carriers, of immense importance for future technological applications, is an external gate, through the electric field effect\cite{2004SciNovoselov} The charge carrier density $n$ is a result of the capacitance of the gate, thus $n$ is linear in the gate voltage $V_g$. Both in chemical doping and in electrical doping, the resulting change in the chemical potential can be estimated by $\delta V={\text{sgn}}(n)\hbar v_f \sqrt{\pi |n|}$.

As a result, the buckling is characterized by ripples with a characteristic size, $\xi=2\pi \sqrt{\frac{\pi (\hbar v_f)^2 \beta \kappa}{4 \ln{2} D \delta V }}$, or numerically:
\begin{equation}
{\xi \approx 144 \Ang \left( \frac{T}{300^\circ{\mathrm{K}}}\right)^{-\frac{1}{2}} \left( \frac{D}{30\ev}\right)^{-\frac{1}{2}} \left( \frac{|n|}{10^{12}{\text{cm}}^{-2}}\right)^{-\frac{1}{4}}}.
\end{equation}
In the case of adsorbed molecules, where carrier concentrations on the order of $10^{13} \text{cm}^{-2}$ have been reached,\cite{2008ZhouPhysRevB} we predict ripples of wavelength as small as $100 \Ang$. For electrical doping, where the carrier concentration is controlled by the gate voltage, the size of the induced ripples is easily calculated to be $\xi \approx 270 \Ang \left(\frac{|V_g|}{10 \mathrm{V}}\right)^{-1/4}$.

This rather amazing outcome, that can and should be checked experimentally,
reflects the possibility of geometric response of the graphene to
external potential. It is clear, that this effect is valid mainly for suspended graphene, as the forces exerted by a substrate would pin the structure of the surface to that of the substrate. In addition, it is important that the graphene would not be covered by a thin layer of water, as suggested in Ref.~27, as this will screen the effect of adsorbants. 
\end{section}
\begin{section} {ADDED RESISTIVITY INDUCED BY THE BUCKLING} 
These ripples, predicted for hole-doped graphene only, will increase the resistivity of the matter,\cite{2007NatMeyer,2008RSPTAKatsnelson} creating an asymmetry in the transport properties of hole and electron doped graphene. Such an asymmetry is also expected due to charged impurities in the substrate,\cite{2007AdamPNAS} and due to a formation of a p-n junction at the contacts.\cite{2008PhysRevBHuard}
In order to estimate the contribution of corrugations to the electrical resistivity, we use the calculations of Ref.~9, where this added resistivity is related to the scattering of electrons on excited flexural phonons within the ripples. For low temperatures ${\text{k}}_B T < 2\pi \frac{\hbar v_{ph}}{\xi} \approx 100 ^{\circ}\text{K}$, phonon excitations are not allowed, thus these will not contribute to the resistivity. For high temperatures, one can approximate the contribution to the resistivity by 
\begin{eqnarray} \nonumber
\delta \rho &\approx& \frac{\hbar}{e^2} \left({\frac{{\text{k}}_B T}{\kappa} \frac{\xi}{2\pi a}}\right)^2 \\ \nonumber &\sim& 200-600 \,\Omega \cdot \left( \frac{T}{300^\circ{\mathrm{K}}}\right)\cdot \left( \frac{|n|}{10^{12}{\text{cm}}^{-2}}\right)^{-1/2}, 
\end{eqnarray}
where $a$ is the interatomic distance.

As previously mentioned, corrugations on the order of $50-150\Ang$ were found in suspended graphene sheets.\cite{2007NatMeyer} This fact can be explained by the effect presented here, through molecular hole-doping. However, the hole-doped buckling can not reproduce the buckling found in quantum Monte-Carlo simulations, which do not take into account the $\pi$-electrons.\cite{2007NatMaFasolino} In Refs.~16 and 17, these ripples were associated with quenched-ripples originating in a substrate, or with peculiarities in the carbon bond. The production of ripples in the process of isolation of one graphene layer, which competes with the effect presented here, can be minimized by depositing graphene on liquid substrates.\cite{2008PhRvLMorozov} However, buckling which originates in the carbon bond will manifest itself, in the elastic free energy, with the same term that leads to the effect discussed here.\cite{2008PhRvBGuinea} As a result it will interfere constructively with the effect predicted here. A distinction between all these effects can be made by the different dependence on the thermodynamic conditions, as the contribution to resistivity of quenched ripples or those originating in the properties of the chemical bond are quadratic in temperature and inverse proportional to the charge carrier concentration\cite{2008PhRvLMorozov} and do not create an electron-hole asymmetry. 

The asymmetry in the electron mobility between hole doped and electron doped samples has been found in many experiments, e.g. Refs.~12 and 13. In particular, recently, Bolotin et al \cite{2008PhRvLBolotin} pointed out an unexpected dependence of the resistivity in charge carrier densities, mainly in hole-doped graphene, that can be qualitatively explained by the effect predicted in the current work.\cite{Gazit}
\end{section}
\begin{section}{SUMMARY} We have presented a theoretical investigation of the effect 2D relativistic fermions have on the structure of the 2D lattice on which they reside. This is achieved by evaluating the contribution of the fermions to the elastic free energy of the surface, a method which leads to the dependence of the elastic constants on the momentum of the excitations. This phenomenon is interpreted as an attenuation of sound waves excited on the surface. In addition, we predict a possibility of structural change reflected in a spontaneous buckling of the surface, due to a non-zero fermion chemical potential.

This study is motivated by, and demonstrated on, graphene. It would be interesting to see whether the same theoretical approach can lead to the better understanding of other materials of two-dimensional character. In graphene, these phenomena are found to have an observable effect on the elastic constants, due to the large deformation energy of the lattice, thus offering a different experimental path to measure this poorly-known quantity. The spontaneous buckling of hole doped graphene, which can be realized by chemical doping or by electrical doping via the electric-field effect, is expected to create ripples which can be directly observed, and to have an effect on the resistivity of this phase of the material. Clearly, an experiment which will provide a direct observation of the rippling as well as the predicted sound attenuation, is called for. 

The theoretical procedure presented here provides an important tool for characterizing the structure of graphene and its transport properties -- the key elements in the design and quality control of any future technological application. 
\end{section}
\begin{acknowledgements}
The author thanks Anton Andreev, Dam Son, George Bertsch and Yusuke Nishida for very helpful
discussions. This work was supported by DOE under Grant No. DE-FG02-00ER41132.
\end{acknowledgements}


\end{document}